\documentclass{article}

\PassOptionsToPackage{numbers, compress,sort}{natbib}

\usepackage[preprint]{neurips_2024}

\usepackage[utf8]{inputenc} 
\usepackage[T1]{fontenc}    
\usepackage{hyperref}       
\usepackage{url}            
\usepackage{booktabs}       
\usepackage{amsfonts}       
\usepackage{nicefrac}       
\usepackage{microtype}      
\usepackage{xcolor}         
\usepackage{graphicx} 
\usepackage{makecell}
\usepackage{pifont}
\usepackage{tikz}           
\usepackage{subcaption}

\title{The CAP Principle for LLM Serving: A Survey of Long-Context Large Language Model Serving}

\author{
  Pai Zeng\\
  Huawei Cloud \& SJTU \\
  \And
  Zhenyu Ning\\
  SJTU \\
  \And
  Jieru Zhao\\
  SJTU \\
  \And
  Weihao Cui\\
  SJTU \\
  \AND
  Mengwei Xu \\
  BUPT \\
  \And
  Liwei Guo \\
  UESTC \\
  \And
  Xusheng Chen \\
  Huawei Cloud \\
  \And
  Yizhou Shan \\
  Huawei Cloud \\
}

\newcommand{\ys}[1]{\textcolor{red}{YS: #1}}

\begin{document}

\maketitle

\begin{abstract}

We survey the large language model (LLM) serving area to understand the intricate dynamics between cost-efficiency
and accuracy, which is magnified by the growing need for longer contextual understanding when deploying models at a massive scale.
Our findings reveal that works in this space optimize along three distinct but conflicting goals: improving serving context length (\textbf{C}), improving serving accuracy (\textbf{A}), and improving serving performance (\textbf{P}).
Drawing inspiration from the CAP theorem in databases, we propose a CAP principle for LLM serving, which suggests that any optimization can improve at most two of these three goals simultaneously. 
%
Our survey categorizes existing works within this framework. 
We find the definition and continuity of user-perceived measurement metrics are crucial in determining whether a goal has been met, akin to prior CAP databases in the wild.
We recognize the CAP principle for LLM serving as a guiding principle,
rather than a formal theorem, to inform designers of the inherent and dynamic trade-offs in serving models.
As serving accuracy and performance have been extensively studied,
this survey focuses on works that extend serving context length and address the resulting challenges.

\end{abstract}

\section{Introduction}
\label{sec:intro}

Large language models (LLMs) and their underlying transformer architecture have revolutionized AI
and have become the bedrock of many emerging applications.
The ecosystem around LLM is on an upward spiral towards artificial general intelligence (AGI): the number of new LLMs and their applications skyrocketed, and as of 2024, LLM-based applications already outperform humans across many tasks such as image classification and visual reasoning~\cite{levels-of-agi-2023, stanford-ai-report-2024}.
High-quality models are essential for any realization of AGI,
but it's equally important to deploy and serve models at a massive scale with a reasonably low cost without compromising their accuracy.
The conflict between serving accuracy and serving performance (e.g., tokens per second.) is a hard one,
prompting extensive research in this area~\cite{xu2024survey, zhou2024survey}.
Generally, there is no one-size-fits-all solution in production settings.
Optimizations to improve performance can lead to reduced accuracy and vice versa.
For example, sparsity and quantization are two common techniques that trade accuracy for better performance.

Unfortunately, this conflict between accuracy and performance has been exacerbated recently
by the growing demand for longer contextual understanding when deploying models in practice~\cite{med-gemini}.
This introduces new complexities as the transformer's attention mechanism exhibits
a quadratic increase in resource consumption with longer contexts~\cite{pope2023efficiently}.
Furthermore, LLMs struggle to utilize information from longer contexts effectively~\cite{lost-in-the-middle-2023}.
Essentially, the need for long-context serving breaks the fragile balance between serving accuracy and performance,
and calls for novel system designs.

To explore the complex relationship between accuracy and performance in large-scale model deployments, particularly for handling long contexts,
we conducted an extensive survey of the LLM serving area.
We highlight three key observations after reviewing related literature.
\begin{enumerate}

\item 
First, we find the scope of a serving system has expanded.
It comprises two system layers: a model serving layer and an agent serving layer. 
The model-layer system runs a given LLM model, typically exposing model inference as its northbound APIs~\cite{vLLM-sosp23,tensorrt-llm}. Works at this layer commonly optimize the model structure~\cite{transformer-xl, zhao2023atom}, cache~\cite{vLLM-sosp23,streamingLLM}, scheduling~\cite{orca-osdi22,tetriserve-2024}, etc.
The agent-layer system sits atop the model-layer system and results from emerging LLM-based system applications that leverage LLM-driven workflow to improve a raw LLM model's accuracy and efficiency while handling complex real-world tasks~\cite{compound-ai-blog}.

\item
Second, we find works in this space
optimize along three distinct goals:
improving serving context length (\textbf{C}ontext),
improving serving accuracy (\textbf{A}ccuracy),
and improving serving performance (\textbf{P}erformance).
Specifically, context means the number of tokens in the context window; accuracy means evaluation metrics on certain tasks (e.g., MMLU), and performance means time-to-first-token, tokens per second, price per million tokens, etc.

\item
Finally, we find a trilemma among the above three goals regardless of which layer they are applied to.
We find that any serving optimization can only improve at most two distinct goals.
We also observe progress in one direction does not lead to progress in others.
For example, using positional embedding to extend a model's range does not improve the model’s accuracy beyond the context length~\cite{roformer-su2021},
and using quantization~\cite{zhao2023atom}, pruning~\cite{compression-survey-2023}, and sparsity~\cite{streamingLLM} enable one to serve a model with faster speed but at the cost of potentially lower accuracy.


\end{enumerate}

Based on the above observations and
inspired by the classical CAP theorem in databases~\cite{cap-wiki},
we propose \textbf{\textit{the CAP principle for LLM serving}}, which states that any given LLM serving optimization, regardless of which system layer it is applied to, can improve at most two of the following goals:
\begin{itemize}
    \item \textbf{C}ontext: The length of context effectively processed and perceived by end users.
    \item \textbf{A}ccuracy: The precision of outputs as evaluated by end users, based on specific task metrics.
    \item \textbf{P}erformance: The efficiency of token processing and generation perceived by end users.
\end{itemize}

\begin{figure}[t]
    \centering                                     
    \includegraphics[width=0.7\textwidth]{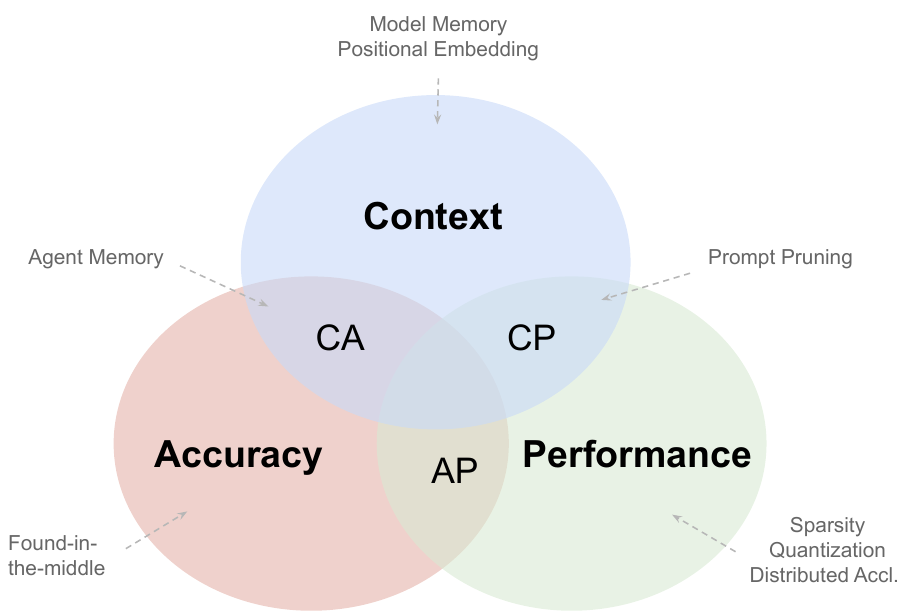}  
    \caption{\textbf{The CAP principle for LLM Serving.}
    C is improving context length,
    A is improving accuracy,
    and P is improving serving performance or cost-efficiency in general.
    It states that any serving optimization can 
    improve at most two of the above three goals.}
    \label{fig:cap}
\end{figure}  

The perspective of the proposed CAP principle emphasizes what \textit{end users} perceive from applying a specific optimization to a remote LLM serving system rather than focusing on a specific component \textit{within} the LLM serving system.
This is crucial because we care whether an LLM serving system as a whole can serve AGI rather than a singular improvement in one direction.
In general, this principle leads to six types of optimizations: C, A, P, CA, CP, and AP, depending on which goals are prioritized.
%
%
%

The LLM's CAP principle is similar to the database's CAP theorem in many ways.
\begin{itemize}

\item 
Both state that you must forfeit at least one goal to achieve the others.
Since our focus is on long-context serving, maintaining a lengthy context (C) is essential.
This leaves us with two options: improving accuracy (A) or improving performance (P).
Improving accuracy relies on devising new algorithms to better leverage the feature of lengthy context.
However, these algorithms could hurt model execution cost-efficiency due to increased FLOPs, hardware-unfriendly operations, etc.
On the other hand, enhancing performance on specific hardware through techniques like quantization and sparsity usually comes at the cost of reduced accuracy. Although there are methods to increase performance without losing accuracy, they generally require additional hardware resources.

\item 
Their goals are measured continuously rather than in binary.
The definition and continuity of user-perceived measurement metrics are crucial in determining whether a goal has been met. Some recent studies have examined this aspect for accuracy~\cite{du2024understanding, schaeffer2024emergent}.
The availability of the database's CAP and the accuracy of the LLM's CAP both range from 0 to 100.
The accuracy of LLM's CAP principle, like the availability of the database's CAP theorem, does not have to be 100\%. It just has to be high enough that end users deem it useful.
Thus, from a system's perspective, an optimization categorized as CP might still be perceived as achieving all three CAP goals if it fulfils the user's accuracy requirements, similar to how CAP is observed in practical databases~\cite{spanner-cap}.

\item 
Both are originally proposed to keep system designers aware of the hard design trade-offs while deploying large-scale systems.

\end{itemize}

We foresee the possibility of a true CAP in the future, in which there is no inherent conflict among these goals.
The proposed CAP principle primarily arises from the use of transformer-based LLMs on existing AI chips, reflecting both the constraints and capabilities of today's hardware and software.
As we progress towards AGI, both models and hardware are expected to evolve significantly.
Emerging technologies are likely to be developed in tandem, with new models specifically designed to optimize performance on the next generation of hardware. 
This synergy between evolving models and hardware is crucial for overcoming current barriers and achieving a true CAP in LLM serving.

\begin{figure}[!b]
    \centering                                     
    \includegraphics[width=0.8\textwidth]{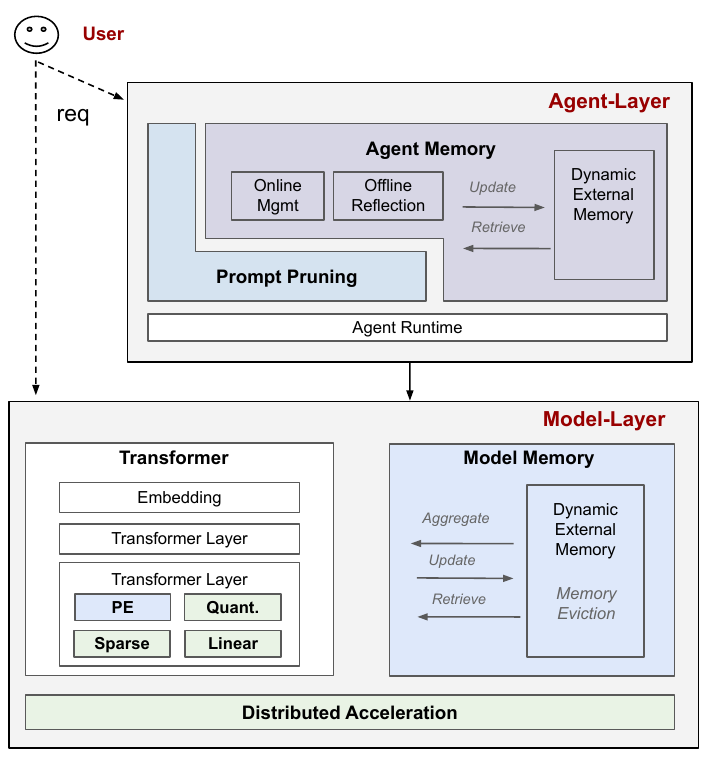}  
    \caption{A modern-day LLM serving system commonly has two layers: a model layer, which runs a given LLM model, and an agent layer, which runs LLM-based system applications.
    PE means Positional Embedding. Quant is short for quantization.
    }
    \label{fig:cap-overview}
\end{figure}

Our survey is organized based on the propose CAP principle.
Compared to prior surveys~\cite{zhou2024survey,xu2024survey,long-context-survey, long-context-survey-dong2023, long-context-survey-wang2024,agent-survey-2023,rag-survey},
we makes two unique contributions.
First, we propose the CAP principle for LLM serving and map existing works onto the CAP landscape to highlight the tension among them. 
Second, we approach the large-scale LLM serving system as a whole rather than focusing on a specific technique (e.g., RAG~\cite{rag-survey}, long-context~\cite{long-context-survey-dong2023}), or a layer (e.g., model~\cite{xu2024survey}, agent~\cite{agent-survey-2023}).
%
%
%
In the rest of the paper, we will discuss works as listed in Table~\ref{tbl:cap-overview} and Figure~\ref{fig:cap-overview}.
We focus on works that extend serving context length and address the resulting accuracy and performance issues.
Specifically, we will cover model memory (Table~\ref{tbl:mm}), positional embedding (Table~\ref{tbl:pe}), found-in-the-middle, distributed acceleration for long context, prompt compression, sparsity (Table~\ref{tbl:sparse}), and agent memory (Table~\ref{tbl:am}).


\section{CAP for LLM Serving}
\begin{table}[h]
    \centering
    \caption{The CAP theorem for LLM serving results in six types.}
    \begin{tabular}{ c | l  }
    \hline
       \textbf{Type} & \textbf{Optimizations} \\
    \hline
       C &  Model Memory, Positional Embedding       \\
    \hline
       A  &  Found-in-the-middle \\
    \hline
       P  &  \makecell{Sparse Attention,  Linear Attention, Distributed Accl., \\  Quantization, Model Pruning}\\
    \hline
       CP  &  Prompt Pruning \\
    \hline
       CA  &  Agent Memory \\
    \hline
       AP  &  N/A \\
    \hline
    \end{tabular}
    \label{tbl:cap-overview}
\end{table}

\subsection{Overview}

We survey the area and map them onto Table~\ref{tbl:cap-overview} across the agent and model layers as depicted in Figure~\ref{fig:cap-overview}.
Remarkably, we can map all existing LLM serving optimization works onto six types resulting from CAP, highlighting that our proposed CAP principle reflects the inherent and long-standing design trade-offs in this area.

An overview of Table~\ref{tbl:cap-overview}:
\begin{itemize}
    \item There are six types: C, A, P, CA, CP, and AP, depending on which goals are
prioritized.
    
    \item \textbf{C}:
    works in this area \textit{only} improve the context length of an LLM serving system.
    Our research identifies two approaches to improve C.
    We dub the first as \textit{Model Memory}, a line of work that augments the transformer with recurrence and dynamic external memory.
    The other is \textit{Positional Embedding}, which extends the context window of the model to a longer context and more tokens.
    
    \item \textbf{A}:
    works in this area address the accuracy issues that arise from long-context serving. A few initial works exist, such as found-in-the-middle, but some forfeit P for a better A.
    
    \item \textbf{P}:
    works in this area improve serving performance or cost-efficiency in general. 
    We focus on two lines of work specifically proposed for improving long-context serving. 
    The first is \textit{distributed acceleration}, which explores sequence
parallelism for faster processing. The second is \textit{sparsity}, which reduces the computation and memory usage for better performance

    
    \item \textbf{CP}: this type of work improves \textit{both} at the same time. We have identified \textit{prompt compression} as this category's only line of work. 

      \item \textbf{CA}: this type of work improves \textit{both} at the same time. We have identified \textit{agent memory} as this category's only line of work.
\end{itemize}

\subsection{Improve Context (C)}
\label{sec:C}

This section surveys work that extends serving systems' context length to address the increasing demand for long-context reasoning. 
We will discuss two approaches.
We dub the first as Model Memory, a line of work that \textit{augments} the transformer architecture with recurrence and dynamic external memory.
The other is Positional Embedding, which \textit{extends} the context window of LLMs to deal with more tokens. 
%

%


\subsubsection{Model Memory}

\begin{table}[t]
    \caption{Comparing model memory works.} 
    \centering 
    \begin{tabular}{ l|ccccc }
    \hline
       \textbf{\thead{Work}} & \textbf{\thead{Memory \\ Aggregation}} & \textbf{\thead{Memory \\ Org.}} & \textbf{\thead{Memory \\ Retrieval}} & \textbf{\thead{Memory \\ Update }} & \textbf{\thead{Memory \\ Eviction}} \\
    \hline
    \hline
        Transformer-XL~\cite{transformer-xl} &  dot-attention &  FIFO & all & None & Discard \\
    \hline
        Compressive Transformer~\cite{compressive} &  dot-attention & FIFO & all & None & Discard \\
    \hline
        Memorizing Transformer~\cite{memorizing-transformers} &  learned-gate & FIFO & kNN & None & Discard \\
    \hline
        Memformer~\cite{memformer} &  dot-attention & random & all & Yes & Yes \\
    \hline
        Memory Transformer~\cite{memory-transformer} &  soft prompt &  random & all & Yes & Yes \\
    \hline
        RMT~\cite{rmt-2022} &  soft prompt & FIFO & all & None & Discard \\
    \hline
        AutoCompressor~\cite{autocompressor} &  soft prompt & FIFO & all & None & Discard \\
    \hline
        Infini-Attention~\cite{infiniattention-google-2024} &  learned gate & random & linear & Yes & Yes \\
    \hline
    \end{tabular}
    \label{tbl:mm}
\end{table}

One way to extend the context length of transformers is by adding memory to hold long-range information.
%
\textit{Model memory} is a term we dubbed for such a line of work, which augments the transformer architecture with recurrence and dynamic external memory. 
At its core, model memory builds a memory system for the transformer, enabling it to examine past long-range information.
%

\paragraph{Taxonomy from the systems perspective}
%
We realize that managing the transformer model's augmented memory is similar to the classical virtual memory management in OS~\cite{legoos-osdi18}, which centers around organizing memory, what to read, update, and what and when to evict.
To this end, we propose to compare model memory works by mapping them into the following five dimensions.
\begin{itemize}
    \item \textbf{Memory Aggregation}: dictates how to aggregate local memory with global memory (retrieved from the augmented memory). It can be attention, learned gate, or soft prompt. 
    
    
    \item \textbf{Memory Organization}: dictates how the external augmented memory is organized. It can be a FIFO buffer or random-access buffer, with fixed memory size. It appears there is no dynamic-sized memory due to capacity concerns.

    \item \textbf{Memory Retrieval}: dictates how and what to retrieve from the augmented memory. Most works will retrieve the full memory, while others retrieve a portion using certain algorithms.
    
    \item \textbf{Memory Update}: dictates how to update the augmented memory when there is a new memory. If it is FIFO memory, the update means enqueue. If it is random memory, the update will use certain algorithms to update all or parts of the memory.
    
    \item \textbf{Memory Eviction}: dictates what to evict when the augmented memory is full. If it is FIFO memory, the eviction discards the tail memory. If it is random memory, no eviction will occur as memory is updated in place.
\end{itemize}

We now delve into works listed in Table~\ref{tbl:mm}.
\begin{itemize}

\item 
Transformer-XL~\cite{transformer-xl} 
adds recurrence to the transformer architecture.
It captures long-term dependency using a per-layer memory buffer
and segments long sequences into fixed-size segments to capture segment-level recurrence between adjacent layers.
%
%
Its memory organization is FIFO, and there are no update rules. 
Old memories are discarded as new segments come in.
During inference, it aggregates the hidden states read from memory and local states from the current segment using dot-product-attention.
Compressive Transformer~\cite{compressive} adds a second-level compressed memory to Transformer-XL. It extends the context further without changing the core mechanisms.
Memorizing Transformer~\cite{memorizing-transformers} takes a slightly different approach.
Instead of reading the whole memory, it uses a kNN algorithm to retrieve from the external memory and aggregates via a learned gate.
The above three works discard information from the distant past.

\item Memformer~\cite{memformer} adds a fixed-size dynamic external memory to the transformer architecture.
It uses random-access memory rather than the FIFO memory used by the former two works.
It segregates the memory into many slots and devises an attention-based algorithm to update the memory slots independently. 
Additionally, it uses a forgetting mechanism to evict memory slots that are not updated for many timestamps.  
By doing so, it attends to \textit{more important} information and claims a theoretically infinite temporal range of memorization.  
%

\item Memory Transformer~\cite{memory-transformer} differs from Memformer~\cite{memformer} in that the former uses soft prompt~\cite{soft-prompt-tuning} to aggregate information from external memory with the current prompt.
It prepends \textit{memory tokens} to tokenized user prompts and uses an unmodified attention module to enable memory tokens to attend to long sequences.


\item
RMT~\cite{rmt-2022} and AutoCompressor~\cite{autocompressor} use soft prompting to add memory tokens to the beginning of the prompts, which is similar to Memory Transformer~\cite{memory-transformer} and segment-level recurrence as in Transformer-XL~\cite{transformer-xl}.
Both are built based on Transformer-XL's code base.
%

\item Infin-Attention~\cite{infiniattention-google-2024}
is the latest work in this category.
It closely integrates compressive and dynamic memory with the vanilla dot-product attention layer to enable models to attend to infinite context length.
It adopts an associative matrix as its memory, allowing random access.
It retrieves memory using linear attention and updates memory using a delta update rule.
It aggregates retrieved memory with a local attention state using a learned gate.
This approach uses less compute and memory compared to the vanilla Transformer-XL.

\end{itemize}

In summary, the model memory line of work augments the original transformer architecture with dynamic and compressive memory,
enabling the model to process long or even infinite contexts.
They differ in how they access the memory, how memory is updated, etc.
Since most of them either discard or compress memory, they inevitably hurt A.
They are neutral in P as they do not address the quadratic complexity in the attention mechanism.

\subsubsection{Positional Embedding}
\begin{table}[b] 
    \centering
    \caption{Comparing positional embedding works.}
    \begin{tabular}{ l|cccc }
    \hline
       \textbf{\thead{Work}} & \textbf{Location} & \textbf{\thead{Require \\ Training}} & \textbf{Adaptive} & \textbf{Integration} \\
    \hline
    \hline
        ALiBi~\cite{alibi} &  After QK multiply &  \checkmark &  \ding{55}  & Add \\
    \hline
        XPOS~\cite{xpos} &  Before QK multiply & \checkmark &  \ding{55} & Multiply  \\
    \hline
        CLEX~\cite{clex} &  Before QK multiply & \checkmark &  \checkmark & Multiply  \\
    \hline
        Linear Interpolation~\cite{linearInterp} &  Before QK multiply & \ding{55} &  \ding{55} & Multiply  \\
    \hline
        NTK Interpolation ~\cite{ntkRope} &  Before QK multiply & \ding{55} &  \ding{55} & Multiply  \\
    \hline
        YaRN~\cite{yarn} &  Before QK multiply &  \checkmark &  \checkmark & Multiply  \\
    \hline
        FIRE~\cite{fire} &  After QK multiply & \checkmark &  \checkmark & Add  \\
    \hline
        LongRoPE~\cite{longrope} &  Before QK multiply & \checkmark &  \checkmark & Multiply  \\
    \hline
    \end{tabular}
    \label{tbl:pe}
\end{table}



This line of work focuses on positional embedding (PE),
enabling LLM to handle long context sequences (hence improving C).
In Table~\ref{tbl:pe}, we compare them across four dimensions.
\begin{itemize}
    \item Location: where is position information being encoded into token representation.
    \item Require training: whether it can plug-and-play without re-training.
    \item Adaptive: whether it can adapt and adjust based on the input.
    \item Integration: how are position representations integrated with token representations.
\end{itemize}

Our discussion below is categorized as extrapolation and interpolation.
\begin{itemize}

\item \textbf{Position Extrapolation}.
This strategy extends the position embedding beyond the max context length used in training.
For example, ALiBi~\cite{alibi} introduces relative positional embedding and a learnable linear bias on attention, which allows the model to dynamically adjust the attention distribution according to the actual length of the sequence.
%
XPOS~\cite{xpos} introduces an additional exponential decay term based on ROPE, which allows attention to decay with increasing relative distance.
CLEX~\cite{clex} models the continuous dynamics as an ordinary differential equation with length scaling factors by generalizing the position-embedding scaling.


\item \textbf{Position Interpolation}:
This strategy scales the input position encoding index range to the context window of a model.
For example, Linear Interpolation~\cite{linearInterp} introduces a position interpolation technique that directly reduces the position index. In this way, the maximum position index matches the previous context window constraints of the pre-training phase and hence extends the context window.
Inspired by Neural Tangent Kernel (NTK) theory, the model using only positional interpolation would have difficulty recognizing the order and position of neighboring tokens. NTK Interpolation~\cite{ntkRope} devised a nonlinear method that changes the base in RoPE to adjust the scaling factor dynamically.
YaRN~\cite{yarn} combines NTK Interpolation and Linear Interpolation and introduces an attention distribution correction strategy to offset the distributional bias in the attention matrix caused by long inputs.
FIRE~\cite{fire} uses a learnable continuous function to map position information to biases and proposes progressive interpolation to address generalization issues when input lengths are outside the training domain.
LongRoPE~\cite{longrope} improves the position interpolation method by recognizing and exploiting non-uniformity in the RoPE dimensions and non-uniformity in the token positions.
PoSE~\cite{pose} introduces a training approach called Positional Skip-wise Method, which emulates extended inputs within a fixed context window by applying tailored skipping bias terms to adjust the position indices for each segment.

\end{itemize}


In summary, the research on positional embeddings enhances the model's ability to generalize positional information
that was not present during the training phase through both extrapolation and interpolation.
These methods vary depending on whether the input position index range is scaled to fit within the model's context window.
They are neutral in C and P. and we believe they are crucial for achieving long-context serving.

\subsection{Improve Accuracy (A)}
\label{sec:A}

A longer C challenges A.
This section focuses on works that address the accuracy issues
that arise from long-context LLM serving.
%
Lost-in-the-middle~\cite{lost-in-the-middle-2023} is a pioneer work
in analyzing how LLMs utilize long context. They found that 
existing LLMs cannot robustly utilize information in a lengthy context,
and the position of the documents will affect the final serving accuracy.
This drawback will limit long-context LLM's usage in practical applications,
leading to biases in outputs.

We find three works to address this issue.
\begin{itemize}
    \item Attention Sorting~\cite{atentionsorting-2024} addresses this issue by placing critical information at the end of the input prompt. They achieve this by performing one step of decoding, sorting documents by the attention they receive (highest attention going last), repeating the process, generate the answer with the newly sorted contexts. Though it could improve A, this approach's limitations are clear: not all tasks map to a set of documents, and the extra sorting adds non-trivial overhead, hurting P.

    \item  Attention Bucket~\cite{AttentionBuckets-2024} uses multiple model replicas, each with a distinct based angle for the rotary position embedding. This creates a unique attention waveform to enhance LLM’s awareness of various contextual positions. This solution works across the model-layer and agent-layer. They improve A but forfeit P because they require multiple replicas to process the input prompt.

    \item 
Found-in-the-middle~\cite{found-in-the-middle-2024} takes a much lighter approach.
They found that the lost-in-the-middle phenomenon likely arises from two factors: casual attention in which LLMs disproportionately favor initial tokens~\cite{streamingLLM} and long-term decay effect of RoPE~\cite{roformer-su2021} that diminishes the attention score of distantly positioned yet semantically meaningful tokens.
Their answer is Multi-scale Positional Encoding (Ms-PoE),
which assigns different scaling ratios to different attention heads to preserve information learned from the pre-training step while using the position indices rescaling to mitigate the long-term decay effect.
This work belongs to the model layer and improves A without adding extra overhead.

\end{itemize}

In summary, improving A under a long C is an area that still needs
close examination. There are some initial works that aim to improve long-context reasoning and understanding, but some of them forfeit P for a better A. We believe more research is needed to improve A and P simultaneously.

\subsection{Improve Performance (P)}
\label{sec:P}

This section covers works that improve P. 
Long-context serving demands significantly more resources in terms of computational flops and memory usage.
From a system's perspective, using principles such as parallelism or approximation to battle these issues is not uncommon.
%
We focus on three lines of work specifically proposed for improving long-context serving: sparse attention, linear attention, and distributed acceleration.

\textbf{Sparse attention} reduces resource usage by selectively focusing on only a subset of the inputs at each attention step.
\textbf{Linear attention} reduces resource usage by approximating the attention calculation through a kernel function that maps the input features into a lower-dimensional space before computing the attention scores. 
Both techniques aim to reduce the quadratic complexity of the traditional attention mechanism. Linear attention does so through dimensionality reduction, while sparse attention uses selective focusing.
Sparse attention is particularly useful when the importance of different parts of the data is non-uniform or when the sequence has a natural locality (like in images or structured text). Linear attention is more suited for tasks where the entire data needs to be compressed and processed efficiently.
\textbf{Distributed acceleration} explores sequence parallelism for faster processing.
We refer readers to ~\cite{zhou2024survey,xu2024survey} for general optimizations that improve P,
for example, paged attention~\cite{vLLM-sosp23}, flash attention~\cite{flashattention}, KV caching~\cite{sglang-2023}, etc.

\subsubsection{Sparse Attention}

This section explores sparsity, a method that enhances computational efficiency by minimizing redundant $QK$ multiplication operations and reducing memory usage. We categorize sparsity techniques into four main types, based on two fundamental aspects.
The first aspect relates to the transformer architecture.
For the Encoder-Decoder architecture, sparsity is applied to selectively ignore less significant interactions between queries and keys in the attention computation. This helps focus computational resources on more crucial elements.
For the Decoder-only architecture, sparsity is used to purge less important data from the key and value cache.
The second aspect focuses on the strategy of identifying which connections between queries and keys are less important. These strategies are divided into two categories including dynamic and static sparsity \cite{salo2}. Dynamic sparsity adapts to the incoming sequence by continually recognizing less important connections between queries and keys and filtering out corresponding tokens at runtime. Static sparsity, on the other hand, uses pre-determined sparse patterns to decide which connections to disregard, simplifying the implementation but potentially sacrificing adaptiveness.

We compare representative sparsity works in Table~\ref{tbl:sparse} across four dimensions.
\begin{itemize}
    \item Sparsity Strategy: whether the sparse pattern of the attention matrix is pre-defined (static) or determined dynamically during inference (dynamic, sometimes also called learned).
    \item Pattern Strategy: composition of retained connections (corresponding to static methods) and technique of obtaining the pattern (corresponding to dynamic methods).
    \item Compensate: whether the system compensates for discarded elements.
    \item Require training: whether a sparsity work plug and play without training.
\end{itemize}

The following discussion is organized based on Table~\ref{tbl:sp-cate}.
\begin{table}[h]
    \caption{The matrix for discussion.}
    \centering
    \begin{tabular}{ c | c |c }
    \hline
       & \textbf{Encoder-Decoder} & \textbf{Decoder-only} \\
    \hline
       \textbf{\makecell{Dynamic \\ Sparsity}} &  (1) & (4)       \\
    \hline
       \textbf{\makecell{Static \\Sparsity}}  &  (2) & (3)\\
    \hline
    \end{tabular}
    \label{tbl:sp-cate}
\end{table}

\textbf{(1) Dynamic Sparsity + Encoder-Decoder.}
In the pre-LLM period, encoder-decoder models dynamically adjust attention patterns at runtime based on input queries and keys, including algorithmic works such as Adaptively Sparse Transformer \cite{adaptivelyTransformer}, Sinkhorn Attention \cite{SinkhornAttention}, Routing transformer \cite{routingTransformer}, Reformer \cite{reformer}, Landmark attention \cite{landmark}, and hardware accelerator works such as $A^3$ \cite{a3}, Spatten \cite{spatten}, Sanger \cite{sanger}, Dota \cite{dota}, Salo2 \cite{salo2}, Acceltran \cite{acceltran}, Fact \cite{Fact}, Energon \cite{Energon} and Dtqatten \cite{DTQAtten}, etc.
These methods filters out irrelevant tokens and generates sparse patterns for crucial attention computation based on input or internal states. They adopt various techniques to determine the sparse pattern at runtime,
such as pruning the attention matrix based on a threshold,
identifying important keys for queries through clustering,
or employing Top-k pruning, among others.
For example, Routing Transformer utilizes clustering to measure the similarity between keys and queries and identifies the Top-k most relevant keys for each query.
Sanger, Acceltran, and Dtqatten derive sparse patterns by masking out elements below a predefined threshold in the approximated score matrix. 

\textbf{(2) Static Sparsity + Encoder-Decoder.}
The quadratic complexity of the attention mechanism incurs heavy computational and memory burdens, especially when the content length is very long.
In scenarios with long input sequences, dynamic sparsity brings about efficieny issues due to the additional overhead of filtering or clustering queries and keys.
This gives rise to static sparsity.
Models like Block-Bert \cite{BlockBert}, Sparse transformer \cite{sparseTransformer}, Longformer \cite{longformer}, BigBird \cite{Bigbird}, Star-transformer \cite{startransformer}, LongT5 \cite{longt5}, LongNet \cite{longnet}, Zebra \cite{zebra} and certain hardware accelerators such as Vitcod \cite{vitcod} and Salo \cite{salo} adopt static sparsity strategies. These works achieve sparsity by constraining attention connections to predefined sparse patterns such as block attention, sliding window attention, global attention, random attention, and dilated attention.
For example, Longformer combines sliding window attention and global attention to capture local and long dependencies, respectively.

\textbf{(3) Static Sparsity + Decoder-Only.}
Now in the LLM period, the model of the decoder-only architecture is becoming mainstream. In the decoding process of the decoder-only transformers, historical keys and values are cached to improve computational efficiency, so sparsity now favors evicting unimportant keys and values in the KV cache. Static sparsity is still applicable on models with decoder-only architecture.
For example, LM-Infinite \cite{han2024lminfinite} and StreamingLLM \cite{streamingLLM} caches the keys and values of the start token and the last $L$ tokens, and only the keys and values in the cache will be used for attention with the current query. 

\textbf{(4) Dynamic Sparsity + Decoder-Only.}
Dynamic sparsity is active again due to the linear complexity of the single-step decoding of the decoder-only architecture. For example, FastGen \cite{ge2024model} chooses the appropriate compression strategy for each attention head in the prefill phase and chooses whether to cache the KV vectors of newly generated tokens according to the compression strategy in the decoding phase. H2O \cite{zhang2023h2o} and Keyformer \cite{adnan2024keyformer} caches the key and value vectors of the last $L$ tokens and the important tokens dynamically selected by attention score at runtime. SparQ Attention \cite{ribar2024sparq} eliminates unimportant key and value vectors based on the approximated attention scores. To compensate for the eliminated value vectors, SparQ Attention additionally maintains the mean value vector of all eliminated value vectors to compute the attention output.
EasyKV \cite{ren2024efficacy} evicts unimportant key and value vectors by a Robust Cache omission policy based on local attention scores and robustness measures.
LESS \cite{dong2024less} uses a low-rank approach based on KV Cache eviction to learn the residual difference between the original attention output and the sparse policy approximation of the attention output, which is accomplished by accumulating the information discarded according to the eviction policy into a constant-sized low-rank cache or state, allowing queries to recover the lost information.
InfLLM \cite{xiao2024infllm} caches keys and values of the start token and the last $L$ tokens and reloads some relevant evicted key and value vectors stored at an external memory by a lookup table. 

\begin{table}[!t]
    \centering 
    \caption{Comparing sparsity works.}
    \begin{tabular}{ l|ccccc }
    \hline
       \textbf{\thead{Work}} & \textbf{\thead{Sparsity \\ Strategy}} & \textbf{\thead{Pattern \\ Strategy}} & \textbf{Compensate} & \textbf{\thead{Require \\ Training}} \\
    \hline
    \hline
        Sparse Transformers~\cite{sparseTransformer} &  Static &  Local + dilated & \ding{55} & \checkmark\\
    \hline
        Adaptively Transformers~\cite{adaptivelyTransformer} &  Dynamic & Topk & \ding{55} & \checkmark\\
    \hline
        Block Attention~\cite{BlockBert} &  Static & Block & \ding{55} & \checkmark\\
    \hline
        ETC~\cite{etc} &  Static & Local + Global & \ding{55} & \checkmark\\
    \hline
        BigBird ~\cite{Bigbird} &  Static & Local + Global + Random & \ding{55} & \checkmark\\
    \hline
        Longformer~\cite{longformer}  &  Static & Local + Global & \ding{55} & \checkmark\\
    \hline
        Reformer~\cite{reformer}    &  Dynamic & LSH & \ding{55} & \checkmark\\
    \hline
        Sinkhorn Attention~\cite{SinkhornAttention}    &  Dynamic & Block + Sort & \ding{55} & \checkmark\\
    \hline
        Routing Transformer~\cite{routingTransformer}    &  Dynamic & Clustering & \ding{55} & \checkmark\\
    \hline
        Star Transformer~\cite{startransformer}       &  Static & Local + Global & \ding{55} & \checkmark\\
    \hline
        LongT5~\cite{longt5}    &  Static & Local + Global & \checkmark & \checkmark\\
    \hline
        LongNet~\cite{longnet}    &  Static & Dilated & \ding{55} & \checkmark\\
    \hline
        Zebra~\cite{zebra}    &  Static & Local or Global & \ding{55} & \checkmark\\
    \hline
        Lankmark Attention~\cite{landmark}    &  Dynamic & Block + Topk & \ding{55} & \checkmark\\
    \hline    
    \hline
        LM-Infinite~\cite{han2024lminfinite}    &  Static & Local + Global & \ding{55} & \checkmark\\
    \hline
        StreamingLLM~\cite{streamingLLM}    &  Static & Local + Global & \ding{55} & \ding{55}\\
    \hline
        H2O~\cite{zhang2023h2o}    &  Dynamic & Local + Topk & \ding{55} & \ding{55}\\
    \hline
        Keyformer~\cite{adnan2024keyformer}    &  Dynamic & Local + Topk & \ding{55} & \ding{55}\\
    \hline
        SparQ Attention~\cite{ribar2024sparq}    &  Dynamic & Topk & \checkmark & \ding{55}\\
    \hline
        EasyKV~\cite{ren2024efficacy}    &  Dynamic & Topk & \ding{55} & \ding{55}\\
    \hline
        LESS~\cite{dong2024less}    &  Dynamic & Topk & \checkmark & \checkmark\\
    \hline
        InfLLM~\cite{xiao2024infllm}    &  Dynamic & Local + Topk & \ding{55} & \ding{55}\\
    \hline
    \end{tabular}
    \label{tbl:sparse}
\end{table}

In summary, 
sparsity improves P by minimizing redundant computation and memory usage.
Most works in this area only improve P but at the cost of lower potentially degraded accuracy.
StreamingLLM~\cite{streamingLLM} is an exception as it achieves both CP by enabling an infinite context window and utilizing efficient attention.
We believe it'd be interesting to explore a combination of model memory, positional embedding, and sparsity optimizations.

\subsubsection{Linear Attention}
    
Linear attention reduces the complexity of the attention mechanism from quadratic to linear with respect to the sequence length.
It does this by approximating the attention calculation through a kernel function that maps the input features into a lower-dimensional space before computing the attention scores.
Specifically,
it replaces the softmax operation with other function, e.g., $sim(Q, K)=\phi(Q)\phi(K)^T$, and computing $\phi(Q)(\phi(K)^TV)$ instead of $sim(Q, K)V$, linear attention reduces quadratic complexity $O(n^2d)$ to linear $O(nrd)$, where $r$ means $\phi()$ maps $\mathbb{R}^d$ to $\mathbb{R}^r$.
Works in this space like Linear Transformer \cite{katharopoulos2020transformers}, Performer \cite{Performer}, and Efficient Attention \cite{Efficient_Attention} define different $\phi()$ to approximate the softmax operation, while Scatterbrain \cite{scatterbrain} and ViTALiTy \cite{vitality} further compensate the low rank function with sparse attentions from Reformer \cite{reformer} and Sanger \cite{sanger} separately. For example, Performer uses positive orthogonal random features (PORF) as a low rank function $\phi()$, while Scatterbrain shows that combining both low-rank linear attention (via function $\phi()$ in Performer) and sparse attention (via locality-sensitive hashing in Reformer) leads to efficient approximation with better performance than individual ones. 

Both linear attention and sparse attention reduce the quadratic complexity of the traditional attention mechanism. Linear attention does so through dimensionality reduction, while sparse attention uses selective focusing.
Both trade A for a better P.

\subsubsection{Distributed Acceleration}

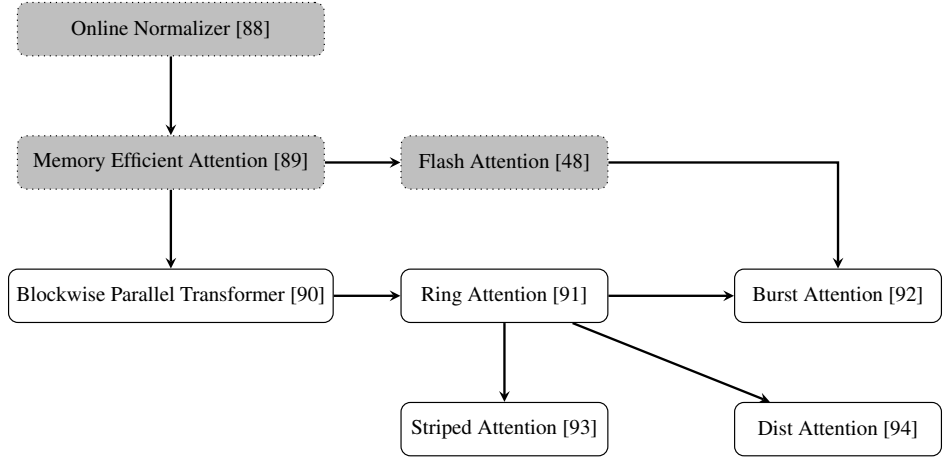
\begin{figure}[t]
    \centering

\tikzstyle{nodestyle_long_gray} = 
[rectangle, rounded corners, minimum width=4.6cm, minimum height=0.8cm, text centered, font=\small, line width=0.5, 
draw=black, 
fill=lightgray, dotted]

\tikzstyle{nodestyle_short_gray} = 
[rectangle, rounded corners, minimum width=3.1cm, minimum height=0.8cm, text centered, font=\small, line width=0.5, 
draw=black, 
fill=lightgray, dotted]

\tikzstyle{nodestyle1} = 
[rectangle, rounded corners, minimum width=4.6cm, minimum height=0.8cm, text centered, font=\small, line width=0.5, 
draw=black, 
fill=white]

\tikzstyle{nodestyle2} = 
[rectangle, rounded corners, minimum width=3.1cm, minimum height=0.8cm, text centered, font=\small, line width=0.5, 
draw=black, 
fill=white]

\tikzstyle{linestyle} = [thick, draw=black, line width=1, ->, >=stealth]

\resizebox{0.9\textwidth}{!}{ 
\begin{tikzpicture}[node distance=2cm]
\node (online) [nodestyle_long_gray] {Online Normalizer~\cite{onlineNomalizer}};
\node (mematten) [nodestyle_long_gray, below of=online] {Memory Efficient Attention~\cite{mem-efficient-attention}};
\node (fa) [nodestyle_short_gray, right of=mematten, xshift=3cm] {Flash Attention~\cite{flashattention}};
\node (bpt) [nodestyle1, below of=mematten] {Blockwise Parallel Transformer~\cite{bpt}};
\node (ring) [nodestyle2, right of=bpt, xshift=3cm] {Ring Attention~\cite{ring-attention}};
\node (burst) [nodestyle2, right of=ring, xshift=3cm] {Burst Attention~\cite{burst-attention}};
\node (stripe) [nodestyle2, below of=ring] {Striped Attention~\cite{striped-attention}};
\node (dist) [nodestyle2, right of=stripe, xshift=3cm] {Dist Attention~\cite{infiniteLLM}};

\draw [linestyle] (bpt) --  (ring);
\draw [linestyle] (mematten) --  (fa);
\draw [linestyle] (mematten) --  (bpt);
\draw [linestyle] (online) --  (mematten);
\draw [linestyle] (ring) --  (burst);
\draw [linestyle] (fa) -|  (burst);
\draw [linestyle] (ring) --  (stripe);
\draw [linestyle] (ring) --  (dist);
\end{tikzpicture} 
}
    
\caption{Works using sequence parallelism. Gray boxes are not tailored for long-context serving.}
\label{fig:dist-accl}
\end{figure}
\begin{figure}[t]
    \centering
    \begin{subfigure}{0.95\textwidth}
        \includegraphics[width=\textwidth]{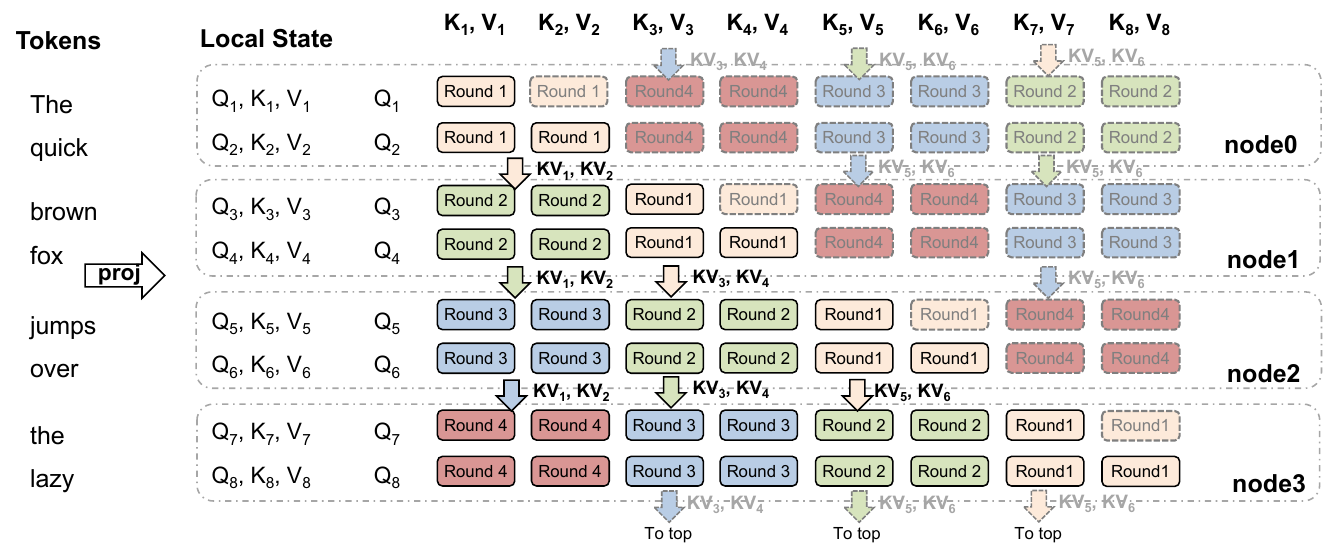}
        \caption{The workflow for computing a single decoder layer in Ring Attention~\cite{ring-attention}. It efficiently implements SP through block-wise computation and overlapping of computation and data transfer.}
        \label{fig:ring}

    \end{subfigure}

    \begin{subfigure}{0.95\textwidth}
        \includegraphics[width=\textwidth]{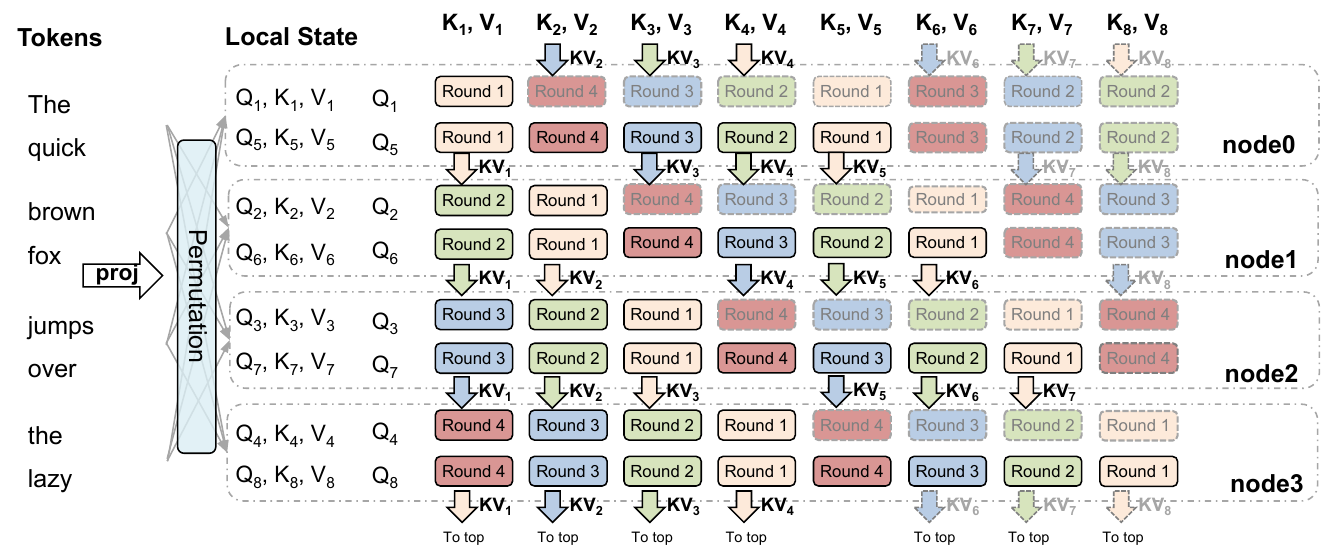}
        \caption{The workflow for computing a single decoder layer in Striped Attention~\cite{striped-attention}. It optimizes Ring Attention by token permutation, which reduces load-imbalance among SP nodes caused by causal masking.}
        \label{fig:striped}
    \end{subfigure}

    \caption{Efficient SP-attention mechanisms used in the prefilling phase of LLM serving.}
\end{figure}

\begin{figure}[!h]
        \includegraphics[width=0.95\textwidth]{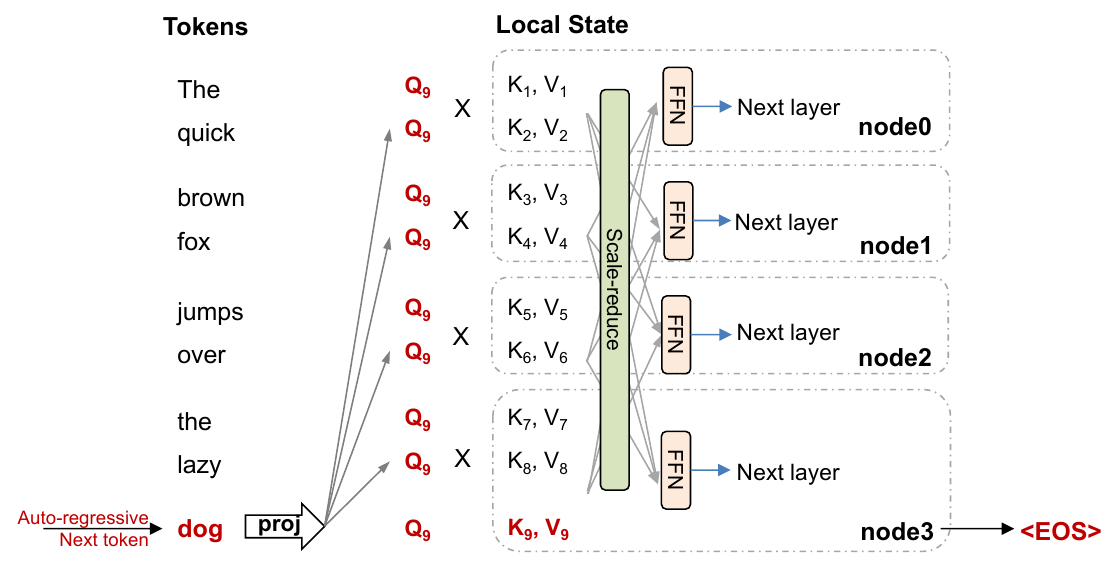}
        \caption{Dist Attention~\cite{infiniteLLM, loongserve}, SP-attention mechanism 
        optimized for the auto-regressive decode phase of LLM serving. In the decode phase, Q length is one and KV is already distributed.}   
    \label{fig:dist}
\end{figure}

We discuss works that explore the Sequence Parallelism (\textbf{SP}) dimension in a distributed fashion.
Here, a long-context inference request is 
segmented into sub-sequences and distributed across nodes for parallel processing. 
While traditional distributed strategies 
like tensor parallelism (TP) or pipeline parallelism (PP) can also enhance inference performance, 
we omit them in this survey because, they are not specifically designed for 
long-context handling and generally serve as orthogonal or complementary to SP optimizations. 


Our analysis unfolds in two steps.
First, we investigate methods to accelerate a single long-context request using SP.
Second, we investigate methods to accelerate a cluster serving long-context requests.



\textbf{Accelerate a Single Request. }
\begin{itemize}

\item
Figure~\ref{fig:dist-accl} shows the relation among this line of research work. 
This line of research can be traced back to the online normalizer work~\cite{onlineNomalizer}, 
a mathematically equivalent method for block-wise softmax calculation that avoids materializing the 
full attention matrix softmax$(QK^T)$.
This method is a foundation for memory-efficient attention~\cite{mem-efficient-attention} and their CUDA implementations~\cite{flashattention, flashattention2}.

\item 
SP was first introduced by Li et al.\cite{SequenceParallelism} 
and has been widely used in distributed LLM training frameworks such as Megatron\cite{megatron-sp}  and Deepspeed~\cite{deepspeed-ul}.
In the context of LLM serving systems, new challenges emerge: 
(1) LLM serving is usually latency-sensitive and thus requires much smaller batch sizes than LLM training;
(2) LLM serving has an auto-regressive decode phase, where the sequence length is only one, but it requires large memory for KV cache storage;
(3) LLM serving usually relies on large fused kernels for improving performance. 
While the feed-forward network (FFN) computations for each token in a sequence are linearly independent, 
the computations for attention are not. 
Consequently, substantial data exchange is involved when computing distributed attention using SP, 
thereby opening significant space for performance optimization.

\item Blockwise Parallel Transfomer (BPT)~\cite{bpt} 
extends this block-wise parallel computation idea from self-attention to a 
fusion of self-attention and FFN. 
BPT computes FFN directly
with each block of Q's attention result without materializing 
the full attention matrix at all, thus reducing memory demands for handling requests with extended contexts.

\item Ring Attention~\cite{ring-attention} is a follow-up 
work of BPT and adapt it for distributed settings. 
As shown in Figure~\ref{fig:ring}, 
it distributes blockwise attention and FFN computations across devices, 
enabling the concurrent communication of key-value blocks in a circular pattern among hosts. 
This setup overlaps communication with the computation of query-key-value blocks and FFN, enhancing efficiency.
Striped Attention~\cite{striped-attention} refines Ring Attention 
by addressing the load imbalance among distributed nodes that 
arises after causal masking, as shown in Figure~\ref{fig:striped}. 
Burst Attention~\cite{burst-attention} enhances Ring Attention by integrating FlashAttention's tiling 
optimizations into per-node computations and incorporating a global optimizer for distributed coordination.
Dist Attention~\cite{infiniteLLM} optimizes Ring Attention specifically 
for the auto-regressive decode phase, as shown in Figure~\ref{fig:dist}
where the query length is just one. 
In the decode phase, Q length is one and KV is already distributed among 
sequence parallel nodes. 

\end{itemize}

\textbf{Accelerate a Cluster}.
\begin{itemize}

\item 
When deploying long-context serving,
the system encounters requests of varying context lengths. 
This diversity poses significant challenges to the LLM serving system, 
the computational and memory requirements for different requests can vary by order of magnitude.
Two concurrent works, Infinite-LLM~\cite{infiniteLLM} and LoongServe~\cite{loongserve}, 
address this challenge using similar ideas: they employ SP to segment requests of different context lengths 
into smaller, manageable pieces and distribute these pieces across the entire cluster for scheduling. 

\item
Infinite-LLM~\cite{infiniteLLM} introduces Dist Attention, an SP attention mechanism optimized 
for the auto-regressive decode phase. Additionally, Infinite-LLM incorporates a global 
memory manager that coordinates the cluster's memory allocation among request pieces, 
taking into account of coherency constraints and fragmentation.

\item
LoongServe~\cite{loongserve}, on the other hand,
proposes Elastic Sequence Parallelism (ESP) to dynamically adjust the degree of parallelism for an inference request with minimal overhead. 
ESP facilitates two optimization strategies: 
(1) reducing the degree of sequence parallelism after the prefill phase and maintaining a 
lower degree of parallelism during the decode phase, as this phase requires less computation 
(per auto-regressive step); 
(2) increasing the degree of sequence parallelism during the auto-regressive phase as the sequence length
grows, which is particularly promising when the LLM is expected to generate long output sequences.

\end{itemize}

In summary, existing works have greatly improved long-context serving's P,
either from a single request's perspective or the cluster's perspective.
We also find potential future directions worth exploring. 
First, although these system works are general to long-context models,
their optimization approaches have no synergy with, or may even contradict the upper-layer model-level optimization.
For instance, attention mechanisms optimized for load balance among SP nodes may perform poorly with context sparsity.
Second, as far as we are concerned, no effort has been made to 
the co-design between the agent-layer techniques and the distributed acceleration systems. 
For instance, after the distributed inference of a request, its 
``memory'' is scattered among multiple nodes, posing challenges for 
the agent system to collect and filter them.
Finally, likewise, no effort examines whether we should and how to
accelerate model memory line of work (see \S\ref{sec:C}) with SP.

\subsection{Improve Context and Performance (CP)}
\label{sec:CP}

This section discusses works that can improve C and P at the same time.
It is challenging to hit two birds with one stone and
we have identified one line of work: prompt compression.


\subsubsection{Prompt Compression}

\begin{table}
    \caption{Comparing prompt compression works.}
    \centering
    \begin{tabular}{ l|c }
    \hline
       \textbf{\thead{Type}} & \textbf{\thead{Work}} \\
    \hline
    \hline
        Block-Box &  \makecell{Selective Context~\cite{selective-context}, LLMLingua~\cite{llmlingua}, 
        \\
        LongLLMLingua~\cite{longllmlingua},
        LLMLingua2~\cite{llmlingua2}}  \\
    \hline
        White-Box & \makecell{Gist-Token~\cite{gisting}, PCCC~\cite{pccc}, \\ ICAE~\cite{icae}, AutoCompressor~\cite{autocompressor}} \\
    \hline
    \end{tabular}
    \label{tbl:pc}
\end{table}

Prompt compression reduces the length of a given prompt while preserving the essential information such that the serving system can process longer context.
Recall that we determine whether C and P have been met based on user-perceived measurement metrics.
We classify this approach as CP because it can shorten the user-provided prompt before being fed into the model, thereby improving user-perceived context length \textit{and} performance.
%
%
We classify works based on whether the LLM model is used as a black box or a while box.

\begin{itemize}
    \item \textbf{Block-box Compression.}
    LLMLingua~\cite{llmlingua} observes significant redundancy in natural languages and proposes a set of methods for compressing prompts by removing tokens. It uses a token-level iterative algorithm to compress prompts. Doing so can preserve the key information within the prompt by considering the conditional dependencies between tokens.
    LongLLMLingua~\cite{longllmlingua} is built based on LLMLingua, adding question-awareness compression by adding contrastive perplexity which captures the distribution shift of tokens w.r.t. questions. 
   LLMLingua2~\cite{llmlingua2} takes a step further, it targets task-agnostic prompt compression. It uses GPT-4 to generate compressed texts from original prompts and a bi-class classifier to drop unneeded tokens.

   \item \textbf{White-box Compression.} This line of work will modify the model architecture in certain ways to achieve compression.
   And they feed the compressed prompts via soft prompting~\cite{soft-prompt-tuning}.
   Gist tokens~\cite{gisting} modifies the transformer attention masks to enable an LLM to compress prompts into smaller sets of “gist” tokens which can be cached and reused for compute efficiency, improving both C and P.
   Another work, PCCC~\cite{pccc}, adds a trainable soft prompt weight to an LLM. Their insight is that prompts used to condition a LLM can be approximately represented by a much smaller set of carefully chosen weights. Their goal is to train the soft prompt weights to mimic a fixed hard prompt as closely as possible.
   ICAE~\cite{icae} takes a different approach. It consists of 2 modules: a learnable encoder adapted from the LLM with LoRA for encoding a long context into a small number of memory slots, and a fixed decoder, which is the LLM itself where the memory slots representing the original context are conditioned on to interact with prompts to accomplish various goals.
   Finally, AutoCompressor~\cite{autocompressor}, a work based on the RMT architecture~\cite{rmt-2022}, builds a segment-level summary token to compress prompts.

\end{itemize}

\label{prompt-comp}

In summary, there are various ways to compress prompts.
One can either treat the LLM as a black-box and use a set of methods to compress prompts before being sent to the black-box LLM.
Alternatively, one can also modify the model architecture to achieve effective compression.
Prompt compression improves user-perceived C and P.
   
   
   


\subsection{Improve Context and Accuracy (CA)}

This section discusses works that can improve C and A at the same time.
We have identified one line of work: agent memory, which manages memory at the agent layer.

\subsubsection{Agent Memory}

One way to extend a serving system's context length and performance is
by implicitly managing the memory and prompt within the agent layer.
We dub this approach as \textit{agent memory}.
It belongs to CA because it can create the illusion of infinite context over fixed-context models and reflect on past memory for higher accuracy in future tasks, thereby improving user-perceived C and A.
Agent memory differs from model memory (covered earlier) in that
agent memory manipulates memory and prompts within agents, while model memory
manipulates memory within models.
They are not conflicting solutions, they complement each other. For example, one can run an agent memory work such as  MemGPT~\cite{packer2023memgpt} atop a model memory work such as Infini-Attention~\cite{infiniattention-google-2024}.

\begin{table}
    \caption{Comparing agent memory works.}
    \centering
    \begin{tabular}{ l|cc }
    \hline
       \textbf{\thead{Work}} & \textbf{\thead{Online Memory \\ Management}} & \textbf{\thead{Offline Memory \\ Reflection}}  \\
    \hline
    \hline
        MemWalker~\cite{memwalker} &   \checkmark & / \\
    \hline
        WebGPT~\cite{webgpt} &  \checkmark & / \\
    \hline
        MemGPT~\cite{packer2023memgpt}  & \checkmark & / \\
    \hline
        TheSim~\cite{thesimagent-2023}  & \checkmark & \checkmark \\
    \hline
        ChatDev~\cite{chatdev-2023}  & \checkmark & \checkmark \\
    \hline
        MetaGPT~\cite{hong2023metagpt} & \checkmark & \checkmark \\
    \hline
        Self-Refine~\cite{self-refine}  &  \checkmark & \checkmark \\
    \hline
        Reflexion~\cite{shinn2024reflexion} & \checkmark & \checkmark \\
    \hline
        MLCopilot~\cite{zhang2023mlcopilot}  & \checkmark & \checkmark \\
    \hline
    \end{tabular}
    \label{tbl:am}
\end{table}
We discuss agent memory work across two dimensions.

\begin{itemize}
    \item 
    \textbf{Online Memory Management}: it indicates whether a solution can dynamically construct a prompt being fed to the model in real-time based on the agent's past memory, external knowledge, and current user prompt. It requires mechanisms to fetch relevant information from past memory and mechanisms to construct prompts.
MemWalker~\cite{memwalker}, WebGPT~\cite{webgpt}, and MemGPT~\cite{packer2023memgpt} are seminal works in this space.
In particular, MemGPT provides the illusion of an infinite context atop a fixed-context model. It builds a multi-level hierarchy and a set of mechanisms to swap memory between the current constructed prompt and external past memory.
Hence, it implicitly improves C.

\item
\textbf{Offline Memory Reflection}: it indicates whether a solution can reflect on an agent's past memory to learn experiences, distill knowledge, remove unnecessary sentences, etc. It requires mechanisms to read and write past memory.
Many agent-based applications adopt this mechanism offline to improve serving accuracy for future tasks~\cite{shinn2024reflexion, self-refine}. For example, agents in ChatDev~\cite{chatdev-2023}, Generative Agents~\cite{thesimagent-2023}, and MLCopilot~\cite{zhang2023mlcopilot} regularly reflect, which synthesizes past memories into higher-level knowledge to improve future task accuracy. Combined, agent memory with feature improves C and A.

\end{itemize}

In summary, agent memory has three key features: online memory management and offline memory reflection. The former meets C, and the latter meets A. 
Agent memory comes close to CAP if one adds prompt compression to it.

\section{Conclusion}

We believe it’s equally important to deploy and serve models at a massive scale with a reasonably low cost without compromising its accuracy, in addition to having a high-quality model.
we survey the LLM serving area to understand the intricate dynamics between cost-efficiency and accuracy
with the growing need for long-context serving.
Our findings reveal that works in this space optimize along three distinct
but conflicting goals: improving serving context length (C), improving serving
accuracy (A), and improving serving performance (P).
We propose the CAP principle, which states that any given LLM serving optimization can only improve at most two of the above three goals. 
We closely examine the related literature and find existing works can fall into this category.
Looking forward, we hope this principle is used to inform designers of the inherent and dynamic trade-offs when building large-scale serving systems.

\bibliographystyle{unsrt}
\bibliography{ref}

\end{document}